\documentclass{article}

\usepackage{microtype}
\usepackage{graphicx}
\usepackage{booktabs} %
\usepackage{amsmath}
\usepackage{subcaption}
\usepackage{hyperref}

\usepackage[accepted]{icml2019}

\icmltitlerunning{Almost Unsupervised Text to Speech and Automatic Speech Recognition}

\begin{document}

\twocolumn[
\icmltitle{Almost Unsupervised Text to Speech and Automatic Speech Recognition }

\icmlsetsymbol{equal}{*}

\begin{icmlauthorlist}
	\icmlauthor{Yi Ren}{equal,to}
	\icmlauthor{Xu Tan}{equal,goo}
	\icmlauthor{Tao Qin}{goo}
	\icmlauthor{Sheng Zhao}{fa}
	\icmlauthor{Zhou Zhao}{to}
	\icmlauthor{Tie-Yan Liu}{goo}
\end{icmlauthorlist}

\icmlaffiliation{to}{Zhejiang University}
\icmlaffiliation{goo}{Microsoft Research}
\icmlaffiliation{fa}{Microsoft STC Asia}

\icmlcorrespondingauthor{Tao Qin}{taoqin@microsoft.com}

\icmlkeywords{Machine Learning, ICML}

\vskip 0.3in
]

\printAffiliationsAndNotice{\icmlEqualContribution} %

\begin{abstract}
Text to speech (TTS) and automatic speech recognition (ASR) are two dual tasks in speech processing and both achieve impressive performance thanks to the recent advance in deep learning and large amount of aligned speech and text data. However, the lack of aligned data poses a major practical problem for TTS and ASR on low-resource languages. In this paper, by leveraging the dual nature of the two tasks, we propose an almost unsupervised learning method that only leverages few hundreds of paired data and extra unpaired data for TTS and ASR. Our method consists of the following components: (1) a denoising auto-encoder, which reconstructs speech and text sequences respectively to develop the capability of language modeling both in speech and text domain; (2) dual transformation, where the TTS model transforms the text $y$ into speech $\hat{x}$, and the ASR model leverages the transformed pair $(\hat{x},y)$ for training, and vice versa, to boost the accuracy of the two tasks; (3) bidirectional sequence modeling, which addresses error propagation especially in the long speech and text sequence when training with few paired data; (4) a unified model structure, which combines all the above components for TTS and ASR based on Transformer model. Our method achieves 99.84\% in terms of word level intelligible rate and 2.68 MOS for TTS, and 11.7\% PER for ASR on LJSpeech dataset, by leveraging only 200 paired speech and text data (about 20 minutes audio), together with extra unpaired speech and text data.  
\end{abstract}

\section{Introduction}
\label{intro}
Text to speech (TTS) and automatic speech recognition (ASR) are two popular tasks in speech processing and have attracted a lot of attention in recent years due to advances in deep learning. Nowadays, the state-of-the-art TTS and ASR systems are mostly based on deep neural models and are all data-hungry, which brings challenges on many languages that are scarce of paired speech and text data. Therefore, a variety of techniques for low-resource and zero-resource ASR and TTS have been proposed recently, including unsupervised ASR~\citep{yeh2018unsupervised,chen2018towards,liu2018completely,chen2018almost}, low-resource ASR~\citep{chuangsuwanich2016multilingual,dalmia2018sequence,zhou2018multilingual}, TTS with minimal speaker data~\citep{chen2018sample,DBLP:conf/nips/JiaZWWSRCNPLW18,DBLP:conf/nips/ArikCPPZ18,DBLP:conf/icml/WangSZRBSXJRS18}, and boosting TTS and ASR simultaneously in a speech chain~\citep{tjandra2017listening,DBLP:conf/interspeech/TjandraS018}.

Works focusing on unsupervised ASR~\citep{chen2018towards,liu2018completely} do not leverage additional information from TTS, which is a dual task to ASR and is of great potential to improve the performance of ASR. Besides, unsupervised ASR typically leverages some task specific algorithms that first segment the speech waveform into words or phonemes and align speech with text data at the segment level. However, TTS usually converts the speech waveform into mel-spectrograms~\citep{wang2017tacotron,ping2018deep,shen2018natural,ren2019fastspeech} or MFCC~\citep{arik2017deep} and processes it in the frame-level. Therefore, the algorithm designed for unsupervised ASR cannot be easily applied to TTS. The works trying to synthesize the voice of a certain speaker with few samples leverage large amount labeled speech and text data from other speakers, which is usually regarded as a transfer learning problem but not an unsupervised learning problem~\citep{chen2018sample,DBLP:conf/nips/ArikCPPZ18}. The speech chain proposed in~\citep{tjandra2017listening} relied on two well-trained ASR and TTS models to further boost accuracy with unpaired speech and text data, which is not applicable in the zero- or low-resource setting.

In this paper, inspired by the dual nature of TTS and ASR tasks, we proposed a novel almost unsupervised method for both TTS and ASR, by leveraging a few amount of paired speech and text data and extra unpaired data. Our method consists of the following components:
\begin{itemize}
\item First, we leverage the idea of self-supervised learning for unpaired speech and text data, to build the capability of the language understanding and modeling in both speech and text domains. Specifically, we use denoising auto-encoder~\citep{vincent2008extracting} to reconstruct corrupt speech and text in an encoder-decoder framework. 
\item Second, we use dual transformation, which is in spirit of back-translation~\citep{sennrich2016improving,he2016dual}, to develop the capability of transforming text to speech (TTS) and speech to text (ASR): (1) the TTS model transforms the text $y$ into speech $\hat{x}$, and then the ASR model leverages the transformed pair $(\hat{x}, y)$ for training; (2) the ASR model transforms the speech $x$ into text $\hat{y}$, and then the TTS model leverages the transformed pair $(\hat{y}, x)$ for training. Dual transformation iterates between TTS and ASR, and boosts the accuracy of the two tasks gradually.
\item Third, considering the speech and text sequence are usually longer than other sequence-to-sequence learning tasks such as neural machine translation, they will suffer more from error propagation~\citep{bengio2015scheduled,DBLP:conf/acl/ShenCHHWSL16,wu2018beyond}, which refers to the problem of the right part of the generated sequence being usually worse than the left part, especially in zero- or low-resource setting due to the lack of supervised data. Therefore, based on denoising auto-encoder and dual transformation, we further leverage bidirectional sequence modeling for both text and speech to alleviate the error propagation problem\footnote{We train the models to generate speech and text sequence in both left-to-right and right-to-left directions. During dual transformation, for example, we use the TTS model to transform the text $y$ to the speech $\hat{\overrightarrow{x}}$ (left-to-right) and $\hat{\overleftarrow{x}}$ (right-to-left). Then the ASR model leverages $(\hat{\overrightarrow{x}},y)$ and $(\hat{\overleftarrow{x}}, y)$ for training, where $\hat{\overrightarrow{x}}$ and $\hat{\overleftarrow{x}}$ are of good quality in the left part and right part respectively, preventing the model from being biased to always generate low quality results in the right part.}.
\item At last, we design a unified model structure based on Transformer~\citep{vaswani2017attention} that can take speech or text as input or output, in order to incorporate the above components for TTS and ASR together. 
\end{itemize}

We conduct experiments on the LJSpeech dataset by leveraging only 200 paired speech and text data and extra unpaired data. First, our proposed method can generate intelligible voice with a word level intelligible rate of 99.84\%, compared with nearly 0 intelligible rate if training on only 200 paired data. Second, our method can achieve 2.68 MOS for TTS and 11.7\% PER for ASR, outperforming the baseline model trained on only 200 paired data. Audio samples can be accessed on \url{https://speechresearch.github.io/unsuper/} and we will release the codes soon.

\section{Background}
In this section, we briefly review the background of this work, including sequence-to-sequence learning, and the end-to-end model used for TTS and ASR. 

\subsection{Sequence to Sequence Learning}
We denote the sequence pair $(x,y) \in \mathcal{(X,Y)}$, where $\mathcal{X}$ and $\mathcal{Y}$ are the source and target domain. $x=(x_1,x_2,...,x_m)$ and $y=(y_1,y_2,...,y_n)$, where $m$, $n$ are the lengths of the source and target sequences, and $x_i$ and $y_t$ are the $i$-th and $t$-th element of sequence $x$ and $y$. For example, in ASR, $x$ is a speech sequence, where each element is a frame of mel-spectrogram, if we use mel-spectrum feature to represent a speech waveform, and $y$ is a text sequence, where each element is usually a character or phoneme. A sequence-to-sequence model learns the parameter $\theta$ to estimate the conditional probability $P(y|x;\theta)$, and usually uses log likelihood as the objective function: 
\begin{equation}
\begin{aligned}
\label{equ_s2s_loss}
\mathcal{L}(\theta; \mathcal{(X,Y)}) = \Sigma_{(x,y)\in \mathcal{\mathcal{(X,Y)}}}\log P(y|x;\theta).
\end{aligned}
\end{equation}
The conditional probability $P(y|x;\theta)$ can be further factorized according to the chain rule: $P(y|x;\theta) = \prod_{t=1}^{n} P(y_t | y_{<t}, x; \theta)$, where $y_{<t}$ is the proceeding elements before position $t$. 

Sequence-to-sequence learning~\citep{DBLP:journals/corr/BahdanauCB14,chan2016listen,vaswani2017attention} is developed based on a general encoder-decoder framework: The encoder reads the source sequence and generates a set of representations. After that, the decoder estimates the conditional probability of each target element given the source representations and its preceding elements. The attention mechanism~\citep{DBLP:journals/corr/BahdanauCB14} is further introduced between the encoder and decoder in order to determine which source representation to focus on when predicting the current element, and is  an important component for sequence to sequence learning.

\subsection{TTS and ASR based on the Encoder-Decoder Framework}
TTS and ASR have long been hot research topics in the field of artificial intelligence and are typical sequence-to-sequence learning problems. Recent successes of deep learning methods have pushed TTS and ASR into end-to-end learning, where both tasks can be modeled in an encoder-decoder framework with attention mechanism\footnote{Although some previous works adopt a feed-forward network~\citep{zhang2018deep,yang2018novel} and achieved promising results on ASR, we focus on the encoder-decoder model structure in this work.}. CNN/RNN based models are widely used in TTS and ASR~\citep{wang2017tacotron,shen2018natural,ping2018deep,ping2018clarinet,chan2016listen,chiu2018state}. Recently, Transformer~\citep{vaswani2017attention} has achieved great success and outperformed RNN or CNN based models in many NLP tasks such as neural machine translation and language understanding~\citep{vaswani2017attention,devlin2018bert,li2018close}. Transformer mainly adopts the self-attention mechanism to model  interactions between any two elements in the sequence and is more efficient for sequence modeling than RNN and CNN~\citep{vaswani2017attention}, especially when the sequence is extremely long. Considering the lengths of the speech sequence as well as the character or phoneme sequence are usually long, we use Transformer as the basic encoder-decoder model structure for both TTS and ASR in this paper.

\section{Our Method}
In this section, we first introduce the key components of our method for almost unsupervised learning on TTS and ASR, and then describe the design of our model structure.

\subsection{Denoising Auto-Encoder}
\label{sec_dae}
Given the large amount of unpaired speech and text data, building the capability of representation extraction (how to understand the speech or text sequence) and language modeling (how to model and generate the sequence in the speech and text domain) is the first step for the transformation between speech and text. To this end, we leverage denoising auto-encoder~\citep{vincent2008extracting} to reconstruct the speech and text sequence from the corrupted version of itself. Denoising auto-encoder is a typical way of self-supervised learning and is widely used in unsupervised learning~\citep{artetxe2017unsupervised,lample2017unsupervised,DBLP:conf/emnlp/LampleOCDR18}. The loss function $\mathcal{L}^{dae}$ of the denoising auto-encoder on speech and text data is formulated as follows:
\begin{equation}
\begin{aligned}
\label{equ_dae}
\mathcal{L}^{dae} = & \quad \mathcal{L}_{\mathcal{S}}(x|C(x);\theta^{\mathcal{S}}_{enc}, \theta^{\mathcal{S}}_{dec} ) \\
 & + \mathcal{L}_{\mathcal{T}}(y|C(y); \theta^{\mathcal{T}}_{enc}, \theta^{\mathcal{T}}_{dec}),
\end{aligned}
\end{equation}

where $\mathcal{S}$ and $\mathcal{T}$ denote the set of sequences in the speech and text domain, $\theta^{\mathcal{S}}_{enc}$, $\theta^{\mathcal{S}}_{dec}$, $\theta^{\mathcal{T}}_{enc}$ and $\theta^{\mathcal{T}}_{dec}$ denote the model parameters of the speech encoder, the speech decoder, the text encoder, and the text decoder respectively, $C$ is a corrupt operation that randomly masks some elements with zero vectors, or swaps the elements in a certain window of the speech and text sequences~\citep{lample2017unsupervised}. $\mathcal{L}_{\mathcal{S}}$ and $\mathcal{L}_{\mathcal{T}}$ denote the loss for speech and text target sequence respectively. In general, we have:
\begin{equation}
\mathcal{L}_{\mathcal{S}}(y|x;\theta_{enc}, \theta_{dec}) = \text{MSE}(y,f(x;\theta_{enc}, \theta_{dec})),
\end{equation}
\begin{equation}
\mathcal{L}_{\mathcal{T}}(y|x;\theta_{enc}, \theta_{dec}) = -\log P(y|x;\theta_{enc}, \theta_{dec}))
\end{equation}
where $\text{MSE}$ denotes the mean squared errors for speech.

\subsection{Dual Transformation}
Dual transformation is the key component in leveraging the dual nature of TTS and ASR tasks and develop the capability of transforming text to speech (TTS) and speech to text (ASR). We transform the speech sequence $x$ into text sequence $\hat{y}$ using the ASR model, and then train the TTS model on the transformed pair $(\hat{y}, x)$. Similarly, we train the ASR model on the transformed pair $(\hat{x}, y)$ generated by the TTS model. Dual transformation is in spirit of back-translation~\citep{sennrich2016improving,he2016dual} in neural machine translation, which is one of the most effective ways to leverage monolingual data for translation. The loss $\mathcal{L}^{dt}$ for dual transformation consists of the following two parts:
\begin{equation}
\begin{aligned}
\label{equ_bt}
\mathcal{L}^{dt} =  \mathcal{L}_{\mathcal{S}}(x|\hat{y};\theta^{\mathcal{T}}_{enc}, \theta^{\mathcal{S}}_{dec}) + \mathcal{L}_{\mathcal{T}}(y|\hat{x};\theta^{\mathcal{S}}_{enc}, \theta^{\mathcal{T}}_{dec}),
\end{aligned}
\end{equation}
where $\hat{y} =\arg\max P(y|x;\theta^{\mathcal{S}}_{enc}, \theta^{\mathcal{T}}_{dec} )$ and $\hat{x} =f(y;\theta^{\mathcal{T}}_{enc}, \theta^{\mathcal{S}}_{dec} )$ denote the text and speech sequence transformed from speech $x$ and text $y$ respectively. During model training, dual transformation is running on the fly, where TTS model leverages the newest text sequence transformed by the ASR model for training, and vice versa, to ensure the accuracy of TTS and ASR can gradually improve.

\subsection{Bidirectional Sequence Modeling}
\label{sec_bsm}
Sequence-to-sequence learning usually suffers from error propagation~\citep{bengio2015scheduled,DBLP:conf/acl/ShenCHHWSL16}, which refers to the problem that if an element is mistakenly predicted during inference, the error will be propagated and the future tokens conditioned on this one will be impacted. This will cause accuracy drop that the right part of the generated sequence is worse than the left part. The speech and text sequence\footnote{A speech sequence is usually converted into mel-spectrograms that contain more than hundreds of frames, while a text sequence is usually converted into phoneme sequence and is longer than the original word or sub-word sequence.} are usually longer than the sequence in other NLP tasks such as neural machine translation, and may suffer more from error propagation. For example, we observe in the experiments that during dual transformation, the right part of the generated speech sequence is usually of lower quality than the left part, with repeating words or missing words. As a consequence, the dual task that relies on the transformed data for training will be affected and the right part of the text as well as speech sequence cannot be well-trained. Thus the TTS and ASR model will both be biased to generate low-quality results on the right part of the sequence, especially in the low- or zero-resource setting due to the lack of supervised data. 

In order to solve the above problem, we leverage the bidirectional sequence modeling to generate speech and text sequence in both left-to-right and right-to-left directions. In this way, the right part of the sequence that is always of low quality in the original dual transformation process can be generated in the right-to-left direction with good quality. As a consequence, the dual task that relies on the transformed data for training will benefit from the improved quality on the right part of the sequence, and will be more balanced in the generation quality between the left and right part of the sequence, which will result in higher transformation accuracy than the original left-to-right generation. At the same time, bidirectional sequence modeling can also act as an effect of data augmentation that leverages the data in both directions, which is helpful especially when few paired data are available in the almost unsupervised learning setting. 

We re-formulate the denoising auto-encoder and dual transformation based on bidirectional sequence modeling. The bidirectional denoising auto-encoder can be formulated as follows:
\begin{equation}
\begin{aligned}
\label{eq_dae_bsm}
\mathcal{L}^{\overrightarrow{dae}} = & \quad \mathcal{L}_{\mathcal{S}}(\overrightarrow{x}|C(\overrightarrow{x});\theta^{\mathcal{S}}_{enc}, \theta^{\mathcal{S}}_{dec} ) \\
& + \mathcal{L}_{\mathcal{T}}(\overrightarrow{y}|C(\overrightarrow{y}); \theta^{\mathcal{T}}_{enc}, \theta^{\mathcal{T}}_{dec}), \\
\mathcal{L}^{\overleftarrow{dae}} = & \quad  \mathcal{L}_{\mathcal{S}}(\overleftarrow{x}|C(\overleftarrow{x});\theta^{\mathcal{S}}_{enc}, \theta^{\mathcal{S}}_{dec} ) \\
& + \mathcal{L}_{\mathcal{T}}(\overleftarrow{y}|C(\overleftarrow{y}); \theta^{\mathcal{T}}_{enc}, \theta^{\mathcal{T}}_{dec}),
\end{aligned}
\end{equation}
where we reconstruct the corrupt speech and text sequence in both left-to-right and right-to-left directions, $C(\cdot)$ is the corrupt operation described in Equation~\ref{equ_dae}. We share the model parameters when modeling the sequence in two different directions, which will be described later.

Similarly, the bidirectional dual transformation can be formulated as follows:
\begin{equation}
\begin{aligned}
\small
\label{eq_dt_bsm}
\mathcal{L}^{\overrightarrow{dt}} =& \quad \mathcal{L}_{\mathcal{S}}(\overrightarrow{x}|\hat{\overrightarrow{y}};\theta^{\mathcal{T}}_{enc}, \theta^{\mathcal{S}}_{dec}) \\
& + \mathcal{L}_{\mathcal{S}}(\overrightarrow{x}|R(\hat{\overleftarrow{y}});\theta^{\mathcal{T}}_{enc}, \theta^{\mathcal{S}}_{dec}) \\
& + \mathcal{L}_{\mathcal{T}}(\overrightarrow{y}|\hat{\overrightarrow{x}};\theta^{\mathcal{S}}_{enc}, \theta^{\mathcal{T}}_{dec}) \\
& + \mathcal{L}_{\mathcal{T}}(\overrightarrow{y}|R(\hat{\overleftarrow{x}});\theta^{\mathcal{S}}_{enc}, \theta^{\mathcal{T}}_{dec}), \\
\mathcal{L}^{\overleftarrow{dt}} = 
& \quad \mathcal{L}_{\mathcal{S}}(\overleftarrow{x}|\hat{\overleftarrow{y}};\theta^{\mathcal{T}}_{enc}, \theta^{\mathcal{S}}_{dec}) \\
& + \mathcal{L}_{\mathcal{S}}(\overleftarrow{x}|R(\hat{\overrightarrow{y}});\theta^{\mathcal{T}}_{enc}, \theta^{\mathcal{S}}_{dec}) \\
& + \mathcal{L}_{\mathcal{T}}(\overleftarrow{y}|\hat{\overleftarrow{x}};\theta^{\mathcal{S}}_{enc}, \theta^{\mathcal{T}}_{dec}) \\
& + \mathcal{L}_{\mathcal{T}}(\overleftarrow{y}|R(\hat{\overrightarrow{x}});\theta^{\mathcal{S}}_{enc}, \theta^{\mathcal{T}}_{dec}), 
\end{aligned}
\end{equation}
where $\hat{\overrightarrow{y}} =\arg\max P(\overrightarrow{y}|\overrightarrow{x};\theta^{\mathcal{S}}_{enc}, \theta^{\mathcal{T}}_{dec} )$, 
$\hat{\overleftarrow{y}} = \arg \max \\ P(\overleftarrow{y}|\overleftarrow{x};\theta^{\mathcal{S}}_{enc},\theta^{\mathcal{T}}_{dec})$, $\hat{\overrightarrow{x}} =f(\overrightarrow{y}; \theta^{\mathcal{T}}_{enc}, \theta^{\mathcal{S}}_{dec} )$ and $\hat{\overleftarrow{x}} =f(\overleftarrow{y}; \theta^{\mathcal{T}}_{enc}, \theta^{\mathcal{S}}_{dec} )$
denote the sequence transformed from $x$ and $y$ in both left-to-right and right-to-left directions. $R(\cdot)$ is reverse function that reverses the sequence from left-to-right to right-to-left or the other way around. 
The loss term $ \mathcal{L}_{\mathcal{S}}(\overrightarrow{x}|R(\hat{\overleftarrow{y}});\theta^{\mathcal{T}}_{enc}, \theta^{\mathcal{S}}_{dec})$ and 
$ \mathcal{L}_{\mathcal{T}}(\overrightarrow{y}|R(\hat{\overleftarrow{x}});\theta^{\mathcal{S}}_{enc}, \theta^{\mathcal{T}}_{dec})$ in $\mathcal{L}^{\overrightarrow{dt}}$ can help the model to better learn on the right part of the sequence, which is usually of poor quality due to error propagation. Similar loss terms can be found in $\mathcal{L}^{\overleftarrow{dt}}$.

As shown in Equations~\ref{eq_dae_bsm} and~\ref{eq_dt_bsm}, bidirectional sequence modeling based on denoising auto-encoder and dual transformation shares the models between left-to-right and right-to-left generations, i.e., we can train one model that bidirectionally generates sequence, which can reduce the model parameter. In order to give the model a sense of which direction the sequence will be generated, unlike the conventional decoder using a zero vector as the start element for training and inference, we use two learnable embedding vectors as the two start elements representing the training and inference directions, one from left to right and the other from right to left. Thus we learn four start embeddings in total, two for speech generation and the other two for text generation.

As the speech and text sequence in TTS and ASR are usually monotonously aligned, e.g., the left part of speech sequence in the decoder usually attends to the left part of the text sequence from the encoder in TTS. In order to be consistent with the left-to-right generation, we also feed the source sequence with the right-to-left direction into the encoder when generating the target sequence in the right-to-left direction. Thus we reverse the source sequence to make it consistent with the target sequence, as shown in Equations~\ref{eq_dae_bsm} and ~\ref{eq_dt_bsm}.

\begin{figure*}[!t] 
	\centering
	
	\begin{subfigure}[h]{0.28\textwidth}
		\centering
		\includegraphics[width=\textwidth]{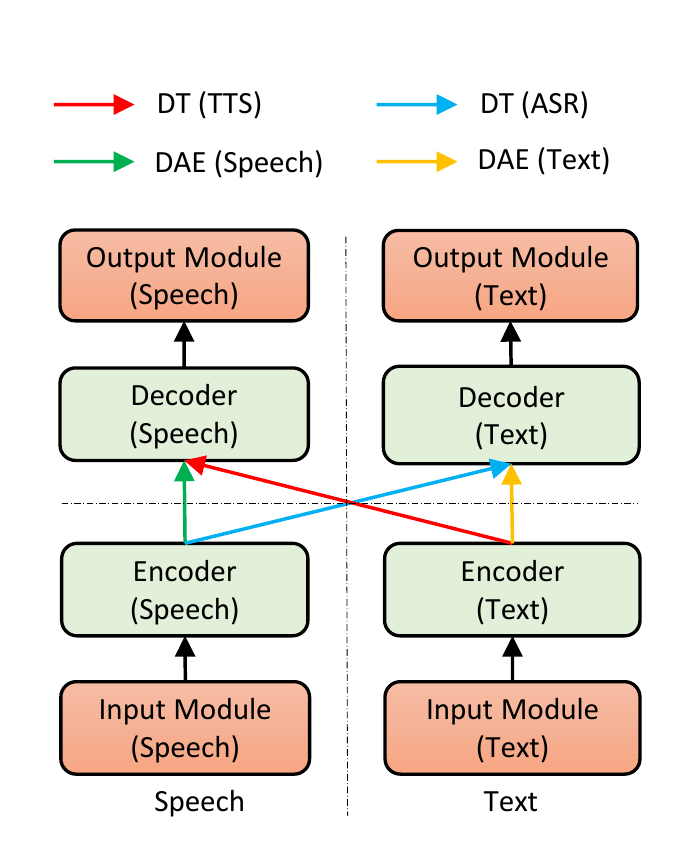}
		\vspace{-0.1cm}
		\caption{Unified training flow}
		\label{fig_archi_overall}
	\end{subfigure}
	\begin{subfigure}[h]{0.35\textwidth}
		\centering
		\includegraphics[width=\textwidth]{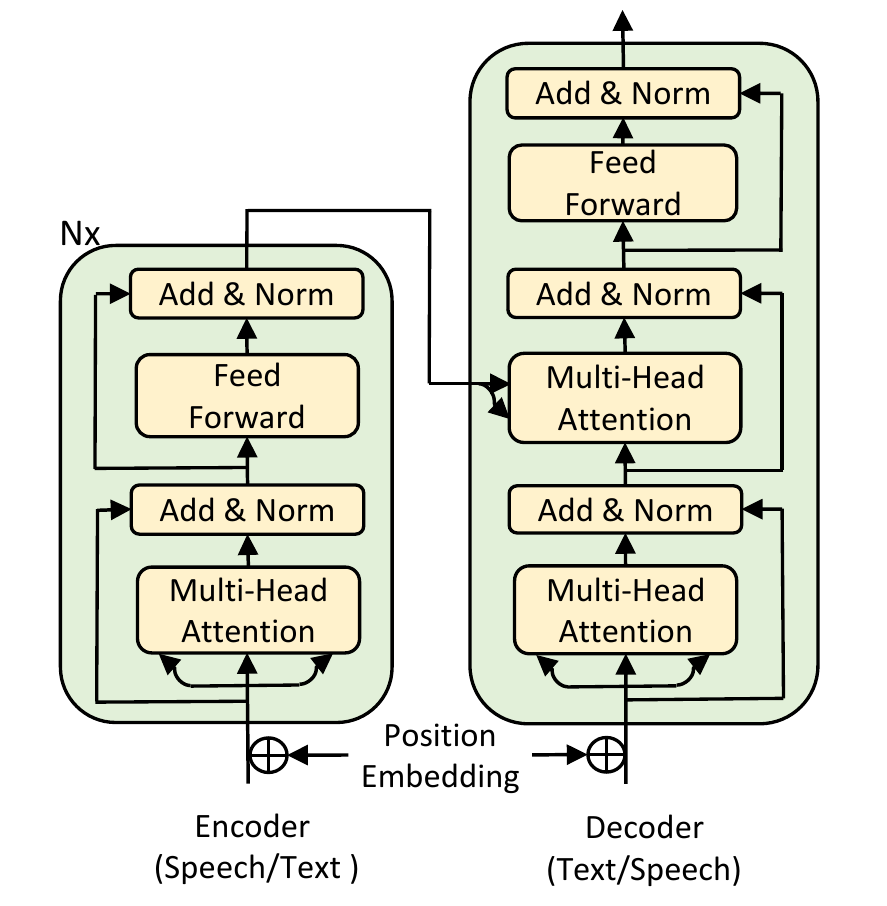}
		\vspace{-0.15cm}
		\caption{Encoder/Decoder for speech/text}
		\label{fig_archi_encdec}
	\end{subfigure}
	\begin{subfigure}[h]{0.29\textwidth}
		\centering
		\includegraphics[width=\textwidth]{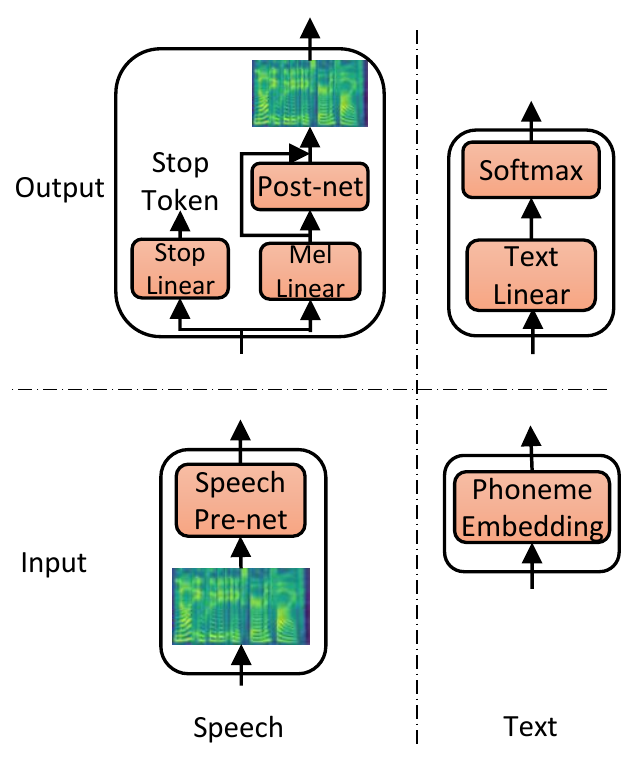}
		\vspace{-0.0cm}
		\caption{Input/Output module for speech/text}
		\label{fig_archi_madality}
	\end{subfigure}
	\caption{The overall model structure for TTS and ASR. Figure (a): The unified training flow of our method, which consists of a denoising auto-encoder (DAE) of speech and text, and dual transformation (DT) of TTS and ASR, both with bidirectional sequence modeling. Figure (b): The speech and text encoder and decoder based on Transformer. Figure (c): The input and output module for speech and text. }
	\label{fig_model_archi}
\end{figure*}

\subsection{Model Structure}
We choose Transformer~\citep{vaswani2017attention} as our basic model, since it has advantages over the conventional RNN/CNN on sequence modeling\footnote{While we choose Transformer as the basic model structure, our method is applicable to other structures such as RNN and CNN.}. The overview of the model structure for TTS and ASR is shown in Figure~\ref{fig_model_archi}. 
We describe the unified training flow of our method as well as the Transformer module and input/output module in this subsection. 

\paragraph{Unified Training Flow}
Figure~\ref{fig_archi_overall} illustrates the unified training flow of our method.  The green and yellow arrows in Figure~\ref{fig_archi_overall} represent the denoising auto-encoder (DAE) for speech and text, while the red and blue arrows represent the dual transformation (DT) from text to speech (TTS) and speech to text (ASR). Both DAE and DT contain the bidirectional sequence modeling as described in Section~\ref{sec_bsm}. 

We also leverage few paired data for bidirectional training, where the loss is as follows:
\begin{equation}
\begin{aligned}
\label{eq_sup_loss}
\mathcal{L}^{\overrightarrow{sup}} = 
& \mathcal{L}_{\mathcal{S}}(\overrightarrow{x}|\overrightarrow{y};\theta^{\mathcal{T}}_{enc}, \theta^{\mathcal{S}}_{dec}, (x, y) \in (\mathcal{S}, \mathcal{T}))\\
& \mathcal{L}_{\mathcal{T}}(\overrightarrow{y}|\overrightarrow{x};\theta^{\mathcal{S}}_{enc}, \theta^{\mathcal{T}}_{dec}, (x, y) \in (\mathcal{S}, \mathcal{T})), \\
\mathcal{L}^{\overleftarrow{sup}} =
& \mathcal{L}_{\mathcal{S}}(\overleftarrow{x}|\overleftarrow{y};\theta^{\mathcal{T}}_{enc}, \theta^{\mathcal{S}}_{dec}, (x, y) \in (\mathcal{S}, \mathcal{T}))\\
& \mathcal{L}_{\mathcal{T}}(\overleftarrow{y}|\overleftarrow{x};\theta^{\mathcal{S}}_{enc}, \theta^{\mathcal{T}}_{dec}, (x, y) \in (\mathcal{S}, \mathcal{T})),
\end{aligned}
\end{equation}
where $(x,y)$ denotes the paired speech and text data.

The total loss of our method is as follows:
\begin{equation}
\begin{aligned}
\label{eq_total_loss}
\mathcal{L} = &\mathcal{L}^{\overrightarrow{dae}} + \mathcal{L}^{\overleftarrow{dae}} + \mathcal{L}^{\overrightarrow{dt}} + \mathcal{L}^{\overleftarrow{dt}} + \mathcal{L}^{\overrightarrow{sup}} + \mathcal{L}^{\overleftarrow{sup}},
\end{aligned}
\end{equation}
where each loss term is described in Equation~\ref{eq_dae_bsm},~\ref{eq_dt_bsm} and~\ref{eq_sup_loss}.

\paragraph{Transformer Module}
The Transformer encoder and decoder for speech and text are shown in Figure~\ref{fig_archi_encdec}. Transformer mainly adopts the self-attention mechanism, which consists of a multi-head attention to extract the cross-position information, and a feed forward network to ensure the nonlinear transformation in each position, each followed by the residual connections and layer normalization. The decoder uses an extra multi-head attention to extract hidden representation from the last layer of the encoder. Both our encoders and decoders are stacked with 4 layers, with the input embedding size, hidden size and feed-forward filter size set to 256, 256 and 1024 respectively. The TTS and ASR share the same model structure for the encoder as well as the decoder, but with different model parameters.

\paragraph{Input/Output Module}
The input and output module for speech and text are shown in Figure~\ref{fig_archi_madality}.
The input module for speech (bottom left in Figure~\ref{fig_archi_madality}) consists of a speech pre-net, which is a 2-layer dense-connected network with hidden size of 256, and the output dimension equals to the hidden size of Transformer. The output module for speech (top left in Figure~\ref{fig_archi_madality}) consists of two components: one is the stop linear layer with the output dimension of 1, plus a sigmoid function to predict if the current decoding step should stop or not, and the other one is a mel linear layer with an additional post-net to generate the mel-spectrogram with 80-dimensional vector in each step. The post-net consists of a 5-layer 1-dimensional convolutional network with hidden size of 256, which aims to refine the quality of the generated mel-spectrograms. We use Griffin-Lim algorithm~\citep{griffin1984signal} to convert the mel-spectrograms into audio\footnote{We can also use WaveNet~\citep{vanwavenet} to generate high quality audio, which we leave to future work.}.

The input module for text (bottom right in Figure~\ref{fig_archi_madality}) is a phoneme embedding, which converts phoneme ID into embedding. We share the parameter of the phoneme embedding with the text linear layer in the output module (top right in Figure~\ref{fig_archi_madality}). The text sequence is first converted into the phoneme sequence with a text-to-phoneme convertor~\citep{sun2019token} before feeding into the model.

\section{Experiments and Results}
In this section, we conduct experiments to evaluate the effectiveness of our proposed method for almost unsupervised TTS and ASR. We first describe the experiment settings, show the results of our method, and conduct some analyses of our method.

\subsection{Training and Evaluation Setup} We choose the speech and text data from LJSpeech dataset~\citep{ljspeech17} for training. LJSpeech contains 13,100 English audio clips and the corresponding transcripts. The total length of the audio is approximate 24 hours. We randomly split the dataset into 3  sets: 12500 samples in training set, 300 samples in validation set and 300 samples in test set. Then we randomly choose 200 audio clips (about 20 minutes) and the corresponding transcripts from the training set as the paired data, and regard the remaining audios and transcripts as unpaired data\footnote{One may argue that there exist implicit aligned signals for the unpaired data that can help the unsupervised training. To verify the implicit aligned signal cannot help, we randomly split the paired data into two half, each half consists of the aligned speech and text data. We train two models of our method, each with different data: the first model is on the speech data from the first half and the text data from the second half, the second model is on the speech and text data both from the first half. We found no difference between the two models in terms of accuracy on TTS and ASR.}. Previous works~\citep{shen2018natural,chan2016listen} convert the transcript text into character sequence as the input or output of the model. In order to alleviate mispronunciation problem, we convert the text sequence into phoneme sequence before feeding into the model, as used in ~\citet{arik2017deep,wang2017tacotron,shen2018natural}. For the speech data, we convert the raw waveform into mel-spectrograms following~\citet{shen2018natural} with 50 ms frame size, 12.5 ms frame hop.

We train the Transformer model on 4 NVIDIA P100 GPUs. The batchsize is 512 sequences in total, which contains 128 sequences for denoising auto-encoder (as shown in Equation~\ref{eq_dae_bsm}, each loss term with 32 sequences) and 256 sequences for dual transformation (as shown in Equation~\ref{eq_dt_bsm}, each loss term with 32 sequences), as well as 128 sequences from the limited paired data (as shown in Equation~\ref{eq_sup_loss}, each loss term with 32 sequences). We upsample the paired data to make it roughly the same with the unpaired data. When training with the denoising auto-encoder loss, we simply mask the elements in the speech and text sequence with a probability of 0.3, as the corrupt operation described in Section~\ref{sec_dae}. We use the Adam optimizer with $\beta_{1}= 0.9$, $\beta_{2} = 0.98$, $\varepsilon = 10^{-9}$ and follow the same learning rate schedule in \citet{vaswani2017attention}. The training takes nearly 3 days.

For evaluation, we mainly use MOS (mean opinion score) for TTS and PER (phoneme error rate) for ASR. For TTS, we also evaluate the intelligibility of the voice~\citep{french1947factors} to verify if we can generate a reasonable speech sequence, considering few paired data are used. We evaluate PER on the test set, from which we further randomly choose 50 paired speech and text data to evaluate the intelligibility rate and mean opinion score (MOS) of different models. We keep the text content consistent among different models so as to exclude other interference factors and only examine audio quality. Each audio is listened by at least 20 testers, who are all native English speakers. 

\subsection{Results}
We first evaluate if our almost unsupervised method can generate audible voice and how much improvement of our method achieves over the baseline system which is trained on 200 paired data only without any other data (denoted as \textit{Pair-200}). \textit{Pair-200} cannot generate any meaningful speech sequence, with nearly 0 intelligible rate. Our method achieves 99.84\% in terms of the word level intelligible rate, which is close to 99.93\%, achieved by the supervised model trained on the whole paired data (denoted as \textit{Supervised}). 

We then compare our method with other systems in terms of MOS on TTS and PER on ASR. The compared systems include: (1) \textit{Pair-200}; (2) \textit{Supervised}; (3) \textit{GT}, the ground truth audio; (4) \textit{GT (Griffin-Lim)}, where we first convert the ground truth audio into mel-spectrograms, and then convert the mel-spectrograms back to audio with Griffin-Lim. Since our method use Griffin-Lim as the vocoder to synthesize audio, we regard the MOS score of \textit{GT (Griffin-Lim)} as the upper bound of the synthesis quality when using Griffin-Lim as the vocoder.

The results are shown in Table~\ref{tab_main_results}. We first compare these systems on the TTS quality in terms of MOS score. The MOS score of the ground truth (\textit{GT}) is 4.54, while 3.21 for \textit{GT (Griffin-Lim)}. It can be seen that there is a big drop when generating audio with Griffin-Lim\footnote{As this work is focusing on almost unsupervised TTS and ASR, we simply use Griffin-Lim as the vocoder in the first step. For future work, we will use advanced techniques for vocoder such as WaveNet~\citep{vanwavenet}.}. \textit{Supervised} that leverages all the paired data can achieve the MOS score of 3.04, which can be regarded as the upper bound of our method. \textit{Pair-200} that leverages only 200 paired data cannot generate any meaningful speech sequence, where we mark its MOS with Null. Our method achieves 2.68 points in terms of MOS, greatly outperforming the \textit{Pair-200} baseline, and is just 0.34 points lower than the \textit{Supervised} that leverages the full paired data. In terms of the accuracy on ASR, our method can achieve 11.7\% PER while \textit{Pair-200} achieves 72.3\% PER, which is much worse than our method. 

\begin{table}[h]
\small
	\centering
	\begin{tabular}{l | c | c}
		\toprule
		Method &  MOS (TTS) & PER (ASR) \\
		\midrule
		\textit{GT} & 4.54 & - \\
		\textit{GT (Griffin-Lim)} & 3.21 &  - \\
		\textit{Supervised} & 3.04  & 2.5\%\\
		\textit{Pair-200} & Null  & 72.3\% \\
		\midrule 
        Our Method & 2.68 & 11.7\%\\
		\bottomrule
	\end{tabular}
	\caption{The comparison between our method and other systems on the performance of TTS and ASR.}
	\label{tab_main_results}
\end{table}

\subsection{Analyses}

\paragraph{Different Components of Our Method}
In order to study the effectiveness of each component of our method, we conduct ablation studies by gradually adding each component to the baseline \textit{Pair-200} system to check the performance changes. We successively add the component: denoising auto-encoder (\textit{DAE}), dual transformation (\textit{DT}) and bidirectional sequence modeling (\textit{BSM}). The results are shown in Table~\ref{tab_ablation_component}. Both \textit{Pair-200} and \textit{Pair-200+DAE} cannot generate reasonable speech sequence, but the PER on ASR is reduced from 72.3\% to 52.0\% after adding \textit{DAE}, mainly due to that \textit{DAE} can leverage more unpaired data than \textit{Pair-200} and build the capability of language modeling both in speech and text domains. When further adding \textit{DT}, the TTS model can generate voice with MOS score of 2.11, and the PER on ASR is reduced from 52.0\% to 15.3\%, which demonstrates the effectiveness of \textit{DT} in boosting the performance from scratch. However, we found that there are many repeating words and missing words in the end of generated speech sequence, due to the error propagation in the long speech and text sequence. Adding \textit{BSM} further brings 0.40 MOS and 3.6\% PER gains, and the repeating words and missing words are greatly reduced, which demonstrates the importance of bidirectional sequence modeling when handling long sequence.
\begin{table}[h]
\small
	\centering
	\begin{tabular}{l | c | c}
		\toprule
		Method &  MOS (TTS) & PER (ASR) \\
		\midrule
		\textit{Pair-200} & Null & 72.3\%  \\
		\textit{Pair-200}+\textit{DAE} & Null & 52.0\% \\
		\textit{Pair-200}+\textit{DAE}+\textit{DT} & 2.11 & 15.3\% \\
		\textit{Pair-200}+\textit{DAE}+\textit{DT}+\textit{BSM} & 2.51 & 11.7\% \\
		\bottomrule
	\end{tabular}
	\caption{Ablation studies on the components of our method.}
	\label{tab_ablation_component}
\end{table}

\begin{figure}[h]
	\centering
	\includegraphics[width=0.40\textwidth]{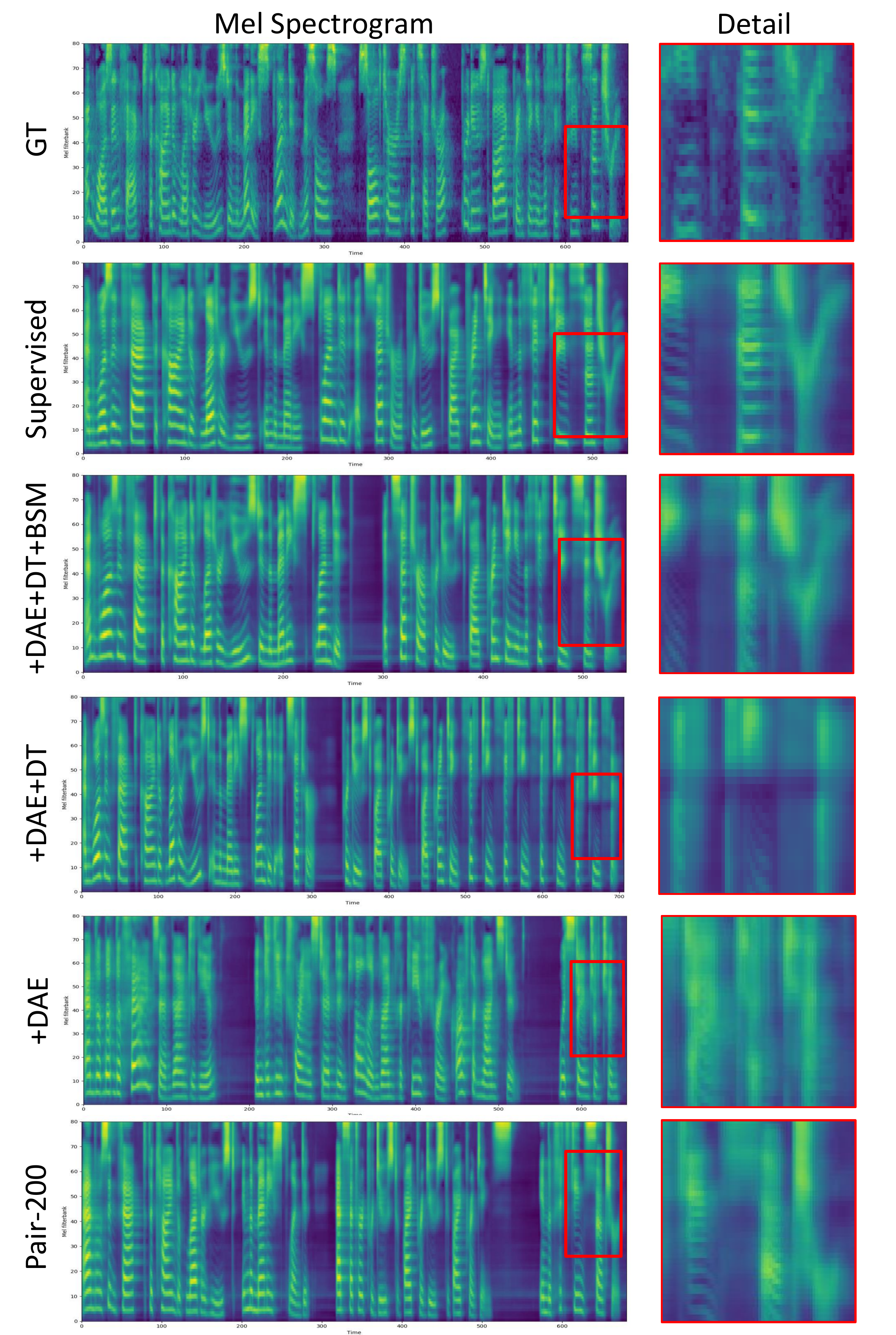}
	\caption{The comparison of mel-spectrograms between different systems. When adding the component of \textit{BSM}, our method can reconstruct the details of the mel-spectrogram in end of the speech sequence. } 
	\label{fig_mel_spectrogram}
\end{figure}

\paragraph{Visualization of Mel-Spectrograms}
We also visualize mel-spectrograms that correspond to the same text in the test set, but generated by different systems, as shown in Figure~\ref{fig_mel_spectrogram}. We mainly compare the systems from Table~\ref{tab_ablation_component}, plus \textit{Supervised} that is trained on the whole paired data, and \textit{GT}, the ground truth mel-spectrograms. Since \textit{Pair-200} and  \textit{Pair-200}+\textit{DAE} cannot generate reasonable speech as shown in Table~\ref{tab_ablation_component}, the details of the mel-spectrograms in the red bounding box are also far different from the ground truth. When adding \textit{DT}, the details of mel-spectrograms are still different from the ground truth, since the red bounding box lies in the end of the mel-spectrogram sequence, and suffers from error propagation. When further adding \textit{BSM}, the details in the bounding box are very close to the ground truth, which also demonstrates the effectiveness of the \textit{BSM} component in our method. If trained on the whole paired data (\textit{Supervised}), the model can reconstruct the details closer to the ground truth. 

\paragraph{Varying Paired Data}
Lastly, we vary the available paired speech and text data for almost unsupervised learning on TTS and ASR, where we choose 100, 200, 300, 400 and 500 data each and verify the performance of our method.  As shown in Table~\ref{tab_ablation_data}, for the PER on ASR, more paired data will consistently improve accuracy. When there are 500 paired data, our method can achieve 4.4\% PER, very close to the PER of the \textit{Supervised} system (2.5\%). For MOS on TTS, our method with 100 paired data cannot generate a meaningful voice. The corresponding PER is also as high as 64.2\%. When gradually adding more paired data, our method can achieve a higher MOS. 

\begin{table}[h]
\small
	\centering
	\begin{tabular}{lcccccc}
		\toprule
		Paired Data & 100 & 200 & 300 & 400 & 500 \\
		\midrule
		PER (ASR) &  64.2\% & 11.7\% & 8.4\% & 5.2\% & 4.4\%   \\
		MOS (TTS) & Null & 2.45 & 2.49 & 2.64 & 2.78 \\
		\bottomrule
	\end{tabular}
	\caption{The PER on ASR with different amount of paired data for our method.}
	\label{tab_ablation_data}
\end{table}

\paragraph{Different Masking Probabilities in DAE}
We further explore different masking probability in the denoising auto-encoder (DAE), which is important to help the model develop the language modeling capabilities in the speech and text domain. We vary different masking probabilities and check the performance of our method in terms of PER on ASR, as shown in Table~\ref{tab_ablation_drop}. Our method can achieve best PER with masking probability of 0.3. 

\begin{table}[h]
\small
	\centering
	\begin{tabular}{lcccccc}
		\toprule
		Probability  & 0.1& 0.2 & 0.3 & 0.4 & 0.5  \\
		\midrule
		PER & 16.1\% &12.9\% & 11.7\% & 12.3\% & 13.4\%  \\
		\bottomrule
	\end{tabular}
	\caption{The PER on ASR with different masking probabilities in the denoising auto-encoder of our method.}
	\label{tab_ablation_drop}
\end{table}

\section{Related Work}
We briefly review related works on TTS and ASR, as well as the zero-/low-resource setting for speech and text.

\paragraph{TTS and ASR} TTS~\citep{arik2017deep,wang2017tacotron,shen2018natural,ping2018clarinet} and ASR~\citep{chiu2018state,zhang2018deep,xiong2018microsoft,yang2018novel} have long been hot research topics in the field of artificial intelligence. Recent successes of deep learning methods have push TTS and ASR into end-to-end modeling. ~\citet{chorowski2014end,chorowski2015attention,chan2016listen,chiu2018state} are the early works to model ASR in an encoder-decoder based framework. In TTS, a variety of methods such as Deep Voice~\citep{arik2017deep}, Tacotron~\citep{wang2017tacotron}, Tacotron2~\citep{shen2018natural}, and ClariNet~\citep{ping2018clarinet} have improved the quality of synthesized speech close to human parity. However, both TTS and ASR require large amounts of high quality paired speech and text data, e.g., hundreds of hours for ASR and dozens of hours for TTS. High quality TTS data for a certain speaker is always hard to collect, and a variety of low-resource languages are lack of paired data, which pose a major practical problem for TTS and ASR.

\paragraph{Zero-/Low-resource TTS and ASR}
Many previous works aim to address the challenge of zero-/low-resource TTS and ASR. 
~\citet{yeh2018unsupervised,chen2018towards,liu2018completely,chen2018almost} tackle the problem of unsupervised ASR and adopt a similar pipeline: speech segmentation, speech embedding learning, speech and text alignment. However, these works just focus on unsupervised ASR without leveraging the dual task (TTS) to improve the accuracy, and usually process the speech in the word or phoneme level, while TTS usually converts the speech waveform into mel-spectrum~\citep{wang2017tacotron} or MFCC~\citep{arik2017deep} and processes in the frame-level. Therefore, the algorithm designed for ASR cannot be easily applied into TTS. ~\citet{chuangsuwanich2016multilingual,dalmia2018sequence,zhou2018multilingual} address the low-resource ASR by formulating the problem in a multilingual scenario, where the data from other languages can act as the effect of data augmentation. ~\citet{chen2018sample,DBLP:conf/nips/JiaZWWSRCNPLW18,DBLP:conf/nips/ArikCPPZ18,DBLP:conf/icml/WangSZRBSXJRS18} synthesize the speech of a target speaker with few paired data, but leverage large amounts of labeled speech and text data from other speakers. Both the two scenarios of low-resource ASR and TTS are typical transfer learning settings and are different from the low-resource setting considered in our work that just leveraging few paired data and extra unlabeled data. The speech chain proposed in~\citep{tjandra2017listening,DBLP:conf/interspeech/TjandraS018} relied on two well-trained ASR and TTS models to further boost the accuracy with unpaired speech and text data, which is not applicable in the zero- or low-resource setting, where it is hard to obtain the ASR and TTS model with good quality. 

Besides the domain of speech processing, some unsupervised learning methods have been proposed in other fields such as neural machine translation. Unsupervised neural machine translation~\citep{artetxe2017unsupervised,lample2017unsupervised,DBLP:conf/emnlp/LampleOCDR18} typically leverage two important components for unsupervised learning: language modeling, which is usually implemented as denoising auto-encoder~\citep{vincent2008extracting} to understand each language in the monolingual context, and back-translation~\citep{sennrich2016improving}, which is one of the most effective ways to leverage monolingual data for translation. 

Our method is carefully designed for the low-resource or almost unsupervised setting with several key components including denoising auto-encoder, dual transformation, bidirectional sequence modeling for TTS and ASR. Some components such as dual transformation are in spirit of the back-translation in neural machine translation. However, different from unsupervised translation where the input and output sequence are in the same domain, speech and text are in different domains and more challenging than translation. We demonstrate in our experiments that our designed components are necessary to develop the capability of speech and text transformation with few paired data. 

\section{Conclusion}
In this work, we have proposed the almost unsupervised method for text to speech and automatic speech recognition, which leverages only few paired speech and text data and extra unpaired data. 
Our method consists of several keys components, including denoising auto-encoder, dual transformation, bidirectional sequence modeling, and a unified model structure to incorporate the above components. We can achieve 99.84\% in terms of word level intelligible rate and 2.68 MOS for TTS, and 11.7\% PER for ASR with just 200 paired data on LJSpeech dataset, demonstrating the effectiveness of our method. The further analyses verify the importance of each component of our method.

For future work, we will push toward the limit of unsupervised learning by purely leveraging unpaired speech and text data, with the help of other pre-training methods~\citep{song2019mass}. We will also leverage an advanced model for the vocoder instead of Griffin-Lim, such as WaveNet, to enhance the quality of the generated audio.

\section*{Acknowledgement}
We thank Jun-Wei Gan, Yi Zhuang from Microsoft STC Asia for the further explorations on this work.

\bibliography{icml2019}
\bibliographystyle{icml2019}

\end{document}